\documentclass[]{elsarticle}
\makeatletter
\def\ps@pprintTitle{%
	\let\@oddhead\@empty
	\let\@evenhead\@empty
	\def\@oddfoot{}%
	\let\@evenfoot\@oddfoot}
\makeatother
\title{Spectroscopic investigations of divertor detachment in TCV}
\usepackage{fullpage,float,multicol}
\usepackage{lineno,hyperref,amsmath,graphicx}
\usepackage{microtype}
\modulolinenumbers[5]
\usepackage{numcompress}
\usepackage{ccicons}

\begin{document}
\setcounter{page}{0}
	\begin{frontmatter}
		
		\title{Spectroscopic investigations of divertor detachment in TCV}

		\author[York,EPFL]{K. Verhaegh\corref{mycorrespondingauthor}}
		\ead{kevin.verhaegh@epfl.ch}
		\cortext[mycorrespondingauthor]{Corresponding author}
		
		\author[York]{B. Lipschultz}
		
		\author[EPFL]{B.P. Duval}
		\author[CCFE]{J.R. Harrison}
		\author[EPFL]{H. Reimerdes}
		\author[EPFL]{C. Theiler}
		
		\author[EPFL]{B. Labit}
		\author[EPFL]{R. Maurizio}
		\author[EPFL]{C. Marini}
		\author[EPFL]{F. Nespoli}
		\author[EPFL]{U. Sheikh}
		\author[SanDiego,EPFL]{C.K. Tsui}
		\author[Padova]{N. Vianello}
		\author[DIFFER]{W.A.J. Vijvers}
		\author{TCV team}
		\author{MST1 team}
		\address[York]{York Plasma Institute, Department of Physics, University of York, Heslington, York, YO10 5DD, United Kingdom}
		\address[EPFL]{Ecole Polytechnique F\'{e}d\'{e}rale de Lausanne (EPFL), Swiss Plasma Center (SPC), CH-1015 Lausanne, Switzerland}
		\address[CCFE]{CCFE, Culham Science Centre, Abingdon, Oxon, OX14 3DB, United Kingdom}
		\address[SanDiego]{University of California San Diego (UCSD), San Diego, CA, USA}
		\address[Padova]{Corsorzio RFX, Corso Stati Uniti 4, 35127 Padova, Italy}
		\address[DIFFER]{FOM Institute DIFFER, 5600 HH Eindhoven, The Netherlands}
		
		\tnotetext[mytitlenote]{\textregistered \hspace{5mm} \vspace{5mm} 2016. This manuscript version is made available under the \href{http://creativecommons.org/licenses/by-nc-nd/4.0/}{CC-BY-NC-ND 4.0} license \ccbyncndeu.}
		
		\begin{abstract}

The aim of this work is to provide an understanding of detachment at TCV with emphasis on analysis of the Balmer line emission. A new Divertor Spectroscopy System has been developed for this purpose. Further development of Balmer line analysis techniques has allowed detailed information to be extracted from the three-body recombination contribution to the n=7 Balmer line intensity.

During density ramps, the plasma at the target detaches as inferred from a drop in ion current to the target. At the same time the Balmer $6\rightarrow2$ and $7\rightarrow2$ line emission near the target is dominated by recombination. As the core density increases further, the density and recombination rate are rising all along the outer leg to the x-point while remaining highest at the target. Even at the highest core densities accessed (Greenwald fraction 0.7) the peaks in recombination and density may have moved not more than a few cm poloidally away from the target which is different to other, higher density tokamaks, where both the peak in recombination and density continue to move towards the x-point as the core density is increased. 


The inferred magnitude of recombination is small compared to the target ion current at the time detachment (particle flux drop) starts at the target. However, recombination may be having more localized effects (to a flux tube) which we cannot discern at this time. Later, at the highest densities achieved, the total recombination does reach levels similar to the particle flux.
 
		\end{abstract}
		
		
	\end{frontmatter}

\section{Introduction}
\label{introduction}

For future fusion devices such as ITER, operating at least in a partially detached state is important for reducing the heat flux incident on the divertor to below engineering limits (10 $\text{MW/m}^2$) \cite{LOARTE2007NF}. Modelling for ITER demonstrates reduction of the peak heat flux near the separatrix by factors of up to 100 due to a number of atomic physics processes including line radiation, charge exchange and recombination \cite{KUKUSHKIN2013JNM}. To address the need for further power removal before exhaust heat reaches the targets, which is needed for a DEMO fusion reactor and beyond, an enhanced understanding of the detachment process would be beneficial, which will enable better models for predicting ITER and DEMO performance and potentially provides insight in enhancing both detachment power and particle loss as well as control of detachment.
 



There has been considerable work utilizing spectroscopic measurements for understanding detachment, where the characteristics of the recombining region are extracted from the Balmer series emission \cite{LIPSCHULTZ1999POP,MCCRACKEN1998NF,MEIGS2013JNM,LOMANOWSKI2015NF,POTZEL2014NF}. Typically a high density recombination front forms at the target and moves rapidly towards the x-point as the core plasma density is increased.

The aim of this study is to develop a detailed understanding of the detachment process at TCV (medium-sized tokamak ($R=0.89 \text{ m}$, $a = 0.25 \text{ m}$, $B_t = 1.4 \text{ T}$)) where low densities should give us insight into how the role of recombination changes as a function of plasma density in the divertor. In addition, this provides an understanding of how detachment on TCV relates to the general experience of detachment. This is required to interpret recent experiments on TCV, which have been performed to investigate how magnetic divertor geometry influences detachment \cite{THEILER2016NF, REIMERDES2016FEC}. 

For this investigation a new spectroscopic diagnostic has been developed for the TCV divertor and improvements for extracting information on recombination and electron temperature from Balmer series spectra have been made. Using spectroscopic measurements we show that the observed high density recombination front at TCV during a density ramp likely stays near the target even after the target ion current drops.



\section{Experimental setup}
\label{ExpSetup}

\subsection{TCV's Divertor Spectroscopy System (DSS)}
\label{DSS}

\begin{figure}[H]
	\centering
	\includegraphics[scale=1.4]{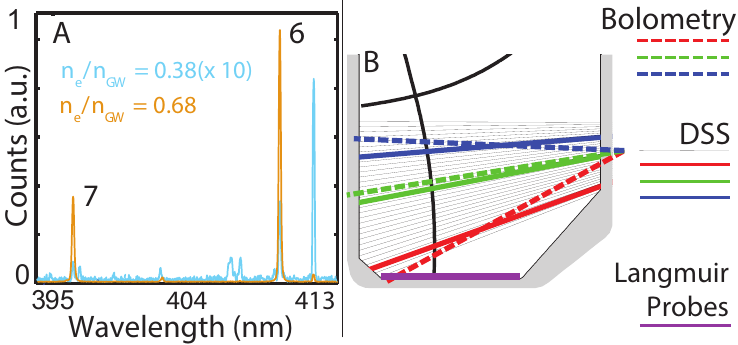}
	\caption{A) Example Balmer line spectra \# 52065, averaged over 100 ms, measured by along the DSS chord closest to the target at two different core densities. B) Primary diagnostic viewing chords used in this work.}
	\label{fig:SpectraFigure}
\end{figure}

The primary measurements of the recombination characteristics are made using a new spectrometer with views of the divertor, which we refer to as the DSS.  The viewing optics provide a poloidal, line-integrated, view of the divertor, yielding 32 lines of sight (figure \ref{fig:SpectraFigure}B). The fibres of each system are coupled to a Princeton Instruments Isoplane SCT 320 spectrometer  coupled to an Andor iXon Ultra 888 EMCCD camera with a 1024 x 1024 pixel sensor. A 1800 l/mm grating was used to allow $n_e$ measurements through Stark broadening of the n=7 Balmer series line with a measured FWHM resolution of 0.06 nm. The system has been absolutely calibrated in intensity ($\sim$15\% inaccuracy) and wavelength ($<0.1$ nm), taking stray light contributions into account.

A dark frame is acquired before and after the plasma discharge, which is subtracted from the measurements. Due to the long frame transfer time (1.2 ms) with respect to the acquisition time (5 - 10 ms), the measurements are susceptible to read-out smear of the CCD \cite{DORRINGTON2005}. At least 90\% of the smearing is removed by post-processing using a numerical matrix-based algorithm.

For the results analysed in this work, the measured spectra have been re-sampled by averaging frames and/or multiple chordal signals over the entire discharge, improving S/N ratio by up to a factor 40 which leads to improved determination of $n_e$ from line fitting (section \ref{NeStark}). 

Figure \ref{fig:SpectraFigure}A shows that the observed intensity of medium-n Balmer lines ($n=6,7$) increases strongly during the density ramp. The observed spectra corresponds to the view line close to the target highlighted in figure \ref{fig:SpectraFigure}B (red), where the locations of the primary diagnostics used in this work are shown.

\subsection{Extracting information on recombination from Balmer lines using a collisional-radiative model}
\label{CollRadMod}

The brightness ($B_{n\rightarrow2}$ in $[\text{photons } \text{m}^{-2} \text{s}^{-1}]$) of a hydrogen Balmer line with quantum number $n$ can be modelled using the Photon Emissivity Coefficients ($PEC_{n\rightarrow2}^{rec,exc}$) [$\text{photons } \text{m }^{3} \text{s }^{-1}$] obtained from the ADAS collisional-radiative model \cite{ADAS} for recombination and excitation, as indicated in equation \ref{eq:EmissEQ}. $B_{n\rightarrow2}$ consists of recombination and excitation parts: $B_{n\rightarrow2}^{rec,exc}$. It is assumed that all line emission comes from a plasma slab with spatially constant electron density $n_e$, electron temperature $T_e$, neutral density $n_o$ and width $\Delta L$.  Additional assumptions are that hydrogen collisional radiative model results are valid for deuterium and that the contribution of charge exchange and molecular reactions (molecular reactions might be significant for detachment in low density plasmas \cite{KUKUSHKIN2016PSI}) to the emission of a certain Balmer line are negligible. For simplicity we have assumed all electrons come from hydrogen ($Z_{eff}=1$), which is discussed in section \ref{impeffect}. 

For further discussion we define $F_{rec}$ as the fraction of total Balmer line radiation due to recombination ($F_{rec} (n) = B_{n \rightarrow 2}^{rec} / B_{n \rightarrow 2}$). We also define $F_{76}$ as the ratio of brightness of the $7\rightarrow2$ and $6\rightarrow2$ Balmer lines ($F_{76} = B_{7\rightarrow2}/B_{6\rightarrow2}$). We define $R_L$ $[\text{rec } / \text{ s} \text{ m}^2]$ as the volumetric recombination rate ($R$ $[\text{rec } / \text{ s} \text{ m}^3]$) line integrated along the line of sight through the plasma for a length $\Delta L$. Although the analysis in this section is mainly focused on the $n = 6,7$ Balmer lines, the analysis strategy is general and can be applied to other Balmer lines. 

\begin{equation}
	B_{n \rightarrow 2} = \underbrace{\Delta L n_e^2 PEC_{n->2}^{rec} (n_e, T_e)}_{B_{n \rightarrow 2}^{rec}} + \underbrace{\Delta L n_o n_e PEC_{n->2}^{exc} (n_e, T_e)}_{B_{n \rightarrow 2}^{exc}}
	\label{eq:EmissEQ}
\end{equation} 

\subsubsection{Using Balmer line ratios to obtain the fraction of Balmer line emission due to recombination}
\label{Frec}

We have developed a method for inferring the recombination contribution to the Balmer line emission, which is important for determining several characteristics of the local plasma. 

\begin{figure}[H]
	\centering
	\includegraphics[scale=1.4]{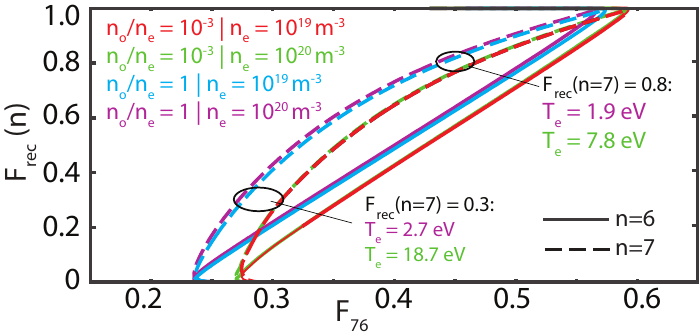}
	\caption{Relation between $F_{76}$ and $F_{rec} (n=6,7)$, in which $T_e$ is varied between 0.2 and 1000 eV, assuming a fixed $n_e = 10^{20} m^{-3}$ and $n_o/n_e = [10^{-3}, 1]$. The value of $T_e$ at $F_{rec} (n=7) = 0.3$ and $F_{rec} (n=7) = 0.8$ is shown for both values of $n_o/n_e$ at $n_e = 10^{20} m^{-3}$.}
	\label{fig:LineRat}
\end{figure}

For a fixed $n_e$ and $n_o$, both $F_{rec} (n)$ and $F_{76}$ only depend on $T_e$. In figure \ref{fig:LineRat} the relation between $F_{rec} (n)$ and $F_{76}$ is shown, where $T_e$ is varied between 0.2 and 1000 eV for each curve, while $n_o/n_e = [10^{-3}, 1]$ and $n_e = 10^{20} \text{ m}^{-3}$ are fixed. Figure \ref{fig:LineRat} indicates the ratio of two Balmer lines (e.g. $F_{76}$) changes as function of $F_{rec}$ and is thus useful to infer the dominance of recombination in the total emission of a particular Balmer line. 

The relation between $F_{rec} (n)$ and $F_{76}$ depends only weakly on $n_e$ and $n_o/n_e$. Divertor pressure measurements with an absolutely calibrated baratron gauge have been used to estimate $n_o$ and indicate $n_o/n_e$ rises from order $10^{-3}$ to order $10^{-1}$ as $T_e$ drops, which is supported by OSM-Eirene modelling \cite{HARRISON2016PSI} and SOLPS-Eirene modelling \cite{WISCHMEIER2005EPFL} of the TCV divertor. Based on the above $n_o/n_e$ estimates, we utilize a $n_o/n_e$ range between 0.01 and 0.25. In this $n_o/n_e$ range and the typical TCV divertor density range (between $10^{19} m^{-3}$ and $10^{20} m^{-3}$) $F_{rec}$ changes by $<0.1$. When inferring $T_e$ and $R_L$ from either $6,7\rightarrow2$ lines (using $F_{rec} (n=6,7)$) the result differs by $<3\%$. Line integration effects (section \ref{ProfEffect}) are negligible to the determination of $F_{rec}$. 


Although $n_o/n_e$ only weakly influences the relation between $F_{rec} (n)$ and $F_{76}$,  it strongly affects the temperature dependence of both $F_{rec} (n)$ and $F_{76}$, as shown in figure \ref{fig:LineRat}. Therefore, to determine $T_e$ from $F_{76}$ an accurate $n_o/n_e$ determination is needed, but is not currently possible. In section \ref{Recom}, we develop another method to derive $T_e$. Note that, even if $n_o/n_e$ would be accurately known, the $T_e$ obtained would be line averaged and weighted over both the excitation and recombination part of the Balmer line emission profile along the line of sight. This is in contrast to the $T_e$ determination described in section \ref{Recom}, where only the recombination part of the Balmer line emission is taken into account.

\subsubsection{Obtaining $R_L$ and $T_e^{avg}$ from absolute Balmer line intensities}
\label{Recom}

We have developed a method for calculating $R_L$, which has the advantage over previous work \cite{TERRY1998POP} that no direct temperature estimate is required in the calculation. The first step is to determine the number of recombinations per photon as in \cite{TERRY1998POP} for a particular Balmer line, which is (assuming the plasma is optically thin) the ratio of the ADAS effective recombination rate coefficient ($ACD (n_e, T_e)$), which takes into account both radiative and three body recombination, and the ADAS $PEC_{n \rightarrow 2}^{rec} (n_e, T_e)$. By multiplying the number of recombinations per emitted photon with $B_{n\rightarrow2}^{rec}$, we obtain $R_L (n_e, T_e, \Delta L)$ [$\text{rec } / \text{m}^2 \text{s}$].

\begin{figure}[H]
	\centering
	\includegraphics[scale=1.4]{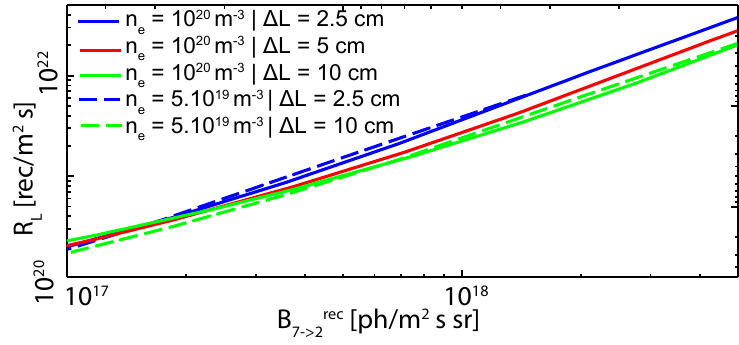}
	\caption{Modelled relation between the $R_L$ and $B_{7\rightarrow2}^{rec}$ for a range of different $n_e$ and $\Delta L$.}
	\label{fig:RecRate}
\end{figure}

Once $F_{rec}$ is determined from figure \ref{fig:LineRat}, we can obtain $B_{n\rightarrow2}^{rec} = F_{rec} (n) \times B_{n\rightarrow2}$, from which we can derive other important characteristics of the plasma along each chord. With fixed $n_e$ and $\Delta L$, both $R_L$ and $B_{n\rightarrow2}^{rec}$ only depend on $T_e$ and a one-to-one relationship between $R_L$ and $B_{n\rightarrow2}^{rec} $ is obtained (figure \ref{fig:RecRate}). In addition, as $T_e$ varies along each curve in figure \ref{fig:RecRate} from 0.2 to 1000 eV, $T_e$ is also obtained when determining $R_L$. We refer to this as $T_e^{avg}$ as it is line averaged and weighted by the recombination part of the Balmer line emission profile along the line of sight. Using $n_e$ (Stark broadening - section \ref{NeStark}), $\Delta L$ and $B_{n\rightarrow2}^{rec}$ both $R_L$ and $T_e^{avg}$ can be determined.

As shown in figure \ref{fig:RecRate}, determining $R_L$ through this method is only weakly affected by $n_e$ and $\Delta L$. The measurement inaccuracy of $R_L$ is generally $\sim$ 40 \% when $F_{rec} \sim 1$ and is mostly due to the inaccuracy in $B_{n\rightarrow2}^{rec}$, which is affected by inaccuracies in both the absolute Balmer line intensity and the Balmer line ratio used to obtain $F_{rec}$. Line integration effects influence $R_L$ by $< 5 \%$, except for cases with a strongly hollow $n_e$ and peaked $T_e$ profile, where $R_L$ can be underestimated by up to 30 \% (section \ref{ProfEffect}). A similar approach as described here could be used to obtain excitation rates and track the excitation region, but with larger uncertainties.

Obtaining $T_e^{avg}$ through the method above has the advantage that less spectral information is needed to obtain $T_e$ than for other methods \cite{LIPSCHULTZ1999POP,TERRY1998POP}. However, this method is sensitive to inaccuracies in $\Delta L$ and is strongly affected by line-integration effects. Assuming peaked $n_e, T_e$ profiles along the line of sight $T_e^{avg}$ is in between 50-100 \% of the peak $T_e$ if $F_{rec} \sim 1$ (section \ref{ProfEffect}). $T_e^{avg}$ should not be used as an absolute $T_e$ measurement, but as an indicator for trends in $T_e$ which shows the role $T_e$ plays in the increase of $R_L$ during a density ramp discharge.

We define $\Delta L$ as the full-width $1/e$ fall-off length of the $n_e$ profile at the target measured by Langmuir probes, which is mapped along the flux surfaces to determine $\Delta L$ for each point where the DSS view line intersects with the separatrix at multiple time points. $\Delta L$ for TCV is generally between 2.5 and 10 cm, depending on the magnetic equilibrium used for that pulse and time. During a density ramp the density profile in the divertor broadens. Together with a constant magnetic equilibrium, $\Delta L$ can increase by up to 70 \%. As a trend in the density profile at the target is not necessarily representable for trends in the density profile across the divertor leg above the target, both the Langmuir probe spatial resolution and experimental variations in $\Delta L$ during a single discharge with constant magnetic equilibrium are used to estimate the uncertainty of $\Delta L$, which makes up at most $ 25 \%$ of the measurement uncertainty in $R_L$.

\subsection{Obtaining $n_e$ from Stark broadening}
\label{NeStark}

The spectrally resolved line profile is affected by Stark broadening. Our chordal measurement provides a density weighted integral of contributions to the line shape and thus of the electron density \cite{LOMANOWSKI2015NF} $(n_e^{Stark})$.

The Stark broadened line shape of a Balmer line can be expressed as a modified Lorentzian \cite{LOMANOWSKI2015NF} as function of $n_e$ and $T_e$, which is a parametrisation of the Microfield Model Method \cite{STEHLE1999AASS}. The spectrometer induces additional instrumental broadening to the emitted spectral line, which is parametrized using a modified asymmetric Lorentzian whose parameters are obtained as function of wavelength and viewing chord.

The experimentally observed Balmer line shape is fitted using a numerical algorithm based on the Gradient Expansion Algorithm \cite{BEVINGTON2003}. The fitting function used is the convolution of Stark broadening, Doppler broadening (depends on $T_i$) \cite{KUNZE2009} and the instrumental line shape. Magnetic effects are neglected. To lower the amount of fitting parameters it is assumed $T_e = T_i = 3 \text{ eV}$. For $T_e$ between 0.6 and 15 eV, the variations in $n_e^{Stark}$ are $<7\%$. For $T_i$ between 0.2 and 15 eV the variations in $n_e^{Stark}$ are $<10\%$. Assuming peaked $n_e$ profiles, $n_e^{\text{Stark}}$ is in between 65 - 100 \% of the peak $n_e$ (section \ref{ProfEffect}). 

The main parameter leading to measurement uncertainty in $n_e^{Stark}$ is the signal/noise level. By fitting synthetic spectra with a level of random noise, we have determined the measurement uncertainty of $n_e^{Stark}$ as function of $n_e$, S/N level and viewing chord. We utilize the $7\rightarrow2$ line for determining $n_e^{Stark}$ since, for the same $n_e$, higher-n Balmer lines lead to wider line shapes, which are more accurately analysed.

\subsection{Investigating line-integration effects on $n_e^{Stark}$, $T_e$ and recombination measurements}
\label{ProfEffect}

The sensitivity of the $n_e^{Stark}$, $T_e^{avg}$, $F_{rec}$ and $R_L$ inferences to line-integration effects have been discussed in sections \ref{Frec}, \ref{Recom} and \ref{NeStark}. These sensitivities have been determined using the methods described in this section.

Line integration effects have been studied by assuming various a priori peaked and hollow $n_e$, $T_e$ profiles along the integration chord. For peaked profiles Gaussian profile shapes have been assumed with widths varying from 0.5 to 7 cm using peak densities: $n_{e,0} = [3, 5, 10] . 10^{19} \text{ m}^{-3}$ and corresponding peak temperatures: $T_{e,0} = [15, 3, 1] \text{ eV}$. A flat neutral density profile using $n_o = [10^{18}, 10^{19}] \text{ m}^{-3}$ has been assumed. 

Using these profiles, the Balmer line emission is modelled at every point of the profile and the corresponding Stark line shape is calculated. The Stark line shapes, weighted by the Balmer line emission, are summed over all points of the profile to obtain a synthetic Balmer line spectrum. $n_e^{Stark}$, $R_L$, $F_{rec}$ and $T_e$ are inferred from the synthetic spectrum using the methods described in sections \ref{CollRadMod} and \ref{NeStark}.

\subsection{The role of impurity concentration on inferred results}
\label{impeffect}

For simplicity in section \ref{CollRadMod} it has been assumed that the hydrogen ion density equals the electron density ($n_H^{+} = n_e$). However, a portion of the electrons can originate from plasma impurities. This can be taken into account by replacing the $n_e^2$ term in $B_{n\rightarrow2}^{rec}$ (equation \ref{eq:EmissEQ}) with $n_e n_H^{+}$, which can be written as $f n_e^2$, where $f=n_H^{+}/n_e$. By including $f$ in equation \ref{eq:EmissEQ} and propagating the effect of $f$ towards the inference of $R_L$, $F_{rec}$ and $T_e^{avg}$ the role of the impurity concentration has been investigated. 

Based on $Z_{eff}$ measurements and fractional abundance modelling through ADAS (using carbon and boron as the main plasma impurity species) we estimate that $f$ is in between 0.6 and 1.0. For this range $F_{rec}$ differs by 0.01, $R_L$ differs by 10 \% and $T_e^{avg}$ differs by 20 \%. Therefore, the impurity concentration is expected to have an effect on the inferred results which is small compared to the estimated uncertainty margins. 


\section{Experimental results}
\label{ExpResults}
In this section we will use the DSS data and analysis techniques described in section \ref{ExpSetup} to illustrate how divertor conditions vary as detachment proceeds in TCV. Connections will be made to other diagnostic measurements to form a more complete picture of the detachment process. Observations of the Balmer line intensity ($B_{n\rightarrow2}$) and the inferred $F_{rec}$ from $F_{76}$ presented in this section correspond to the $n=7$ Balmer line. Our observation is that some of the characteristics of detachment on TCV are similar to that found at other, higher density, tokamaks. However, detachment in TCV does not lead to a large movement of the recombination region.

\subsection{Onset, evolution and dynamics of detachment}
\label{DetachDyn}

\begin{figure}
	\centering
	\includegraphics[scale=1.1]{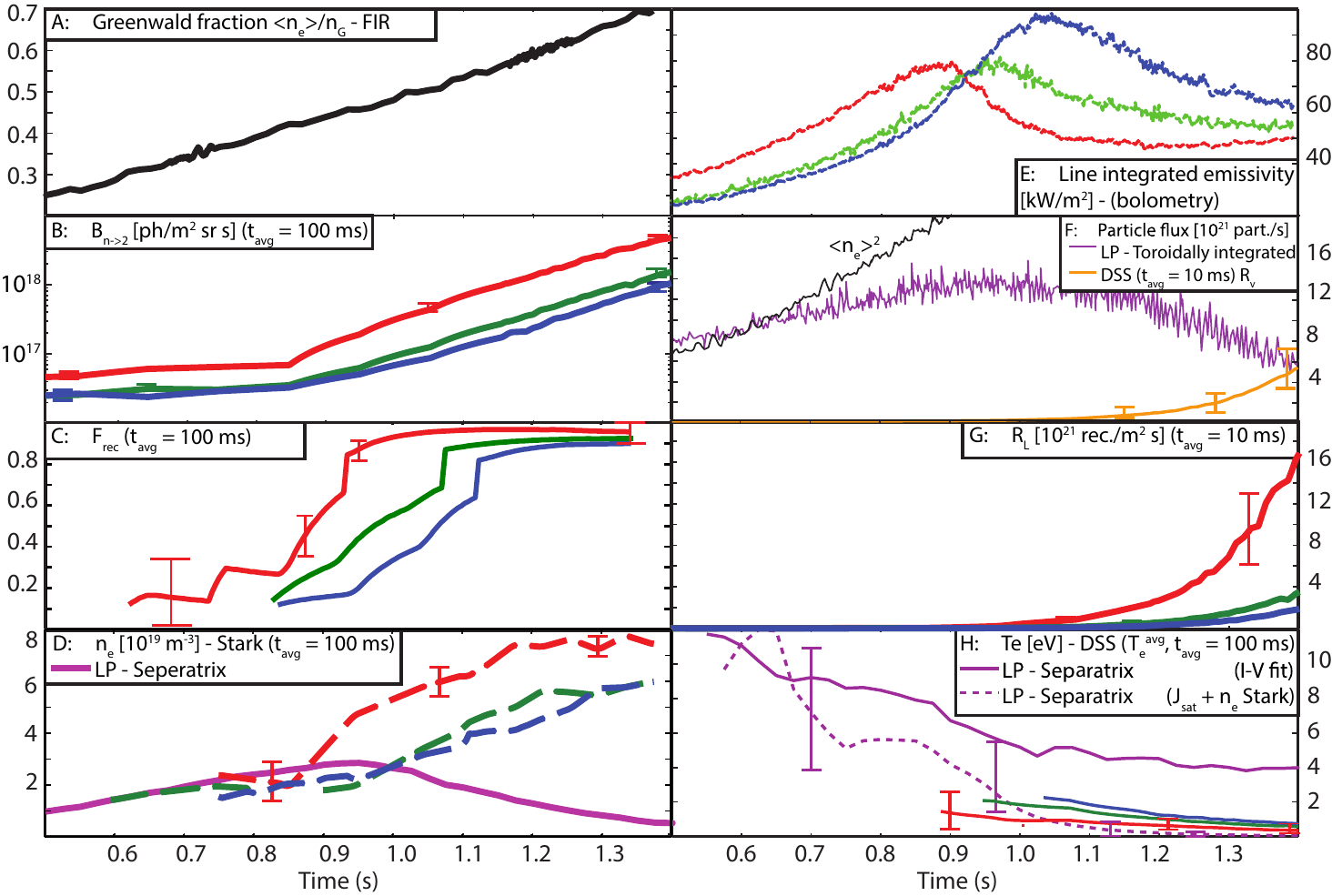}
	\caption{Temporal evolution of several quantities measured by DSS and inferred from DSS measurements for three view lines during a single null density ramp shot (52065). In addition, data obtained from Langmuir probes (LP) and bolometry is shown. The colours of the plot indicate the measurement locations shown in figure \ref{fig:SpectraFigure}B.}
	\label{fig:Overview50648}
\end{figure}

A reference plasma discharge is utilized for illustrating the process of detachment in TCV (\#52065). It has a single null magnetic divertor geometry with a plasma current of 340 kA and a reversed toroidal field direction ($\nabla B$ away from the x-point). The spectroscopic data has been acquired at 200 Hz and has been averaged over a number of time frames to improve S/N level, as indicated in the legends in figure \ref{fig:Overview50648}. The line colour and line style shown in figure \ref{fig:Overview50648}A-H correspond to the diagnostic locations shown in figure \ref{fig:SpectraFigure}B. Similar detachment characteristics as observed for \#52065 have been found for $\sim 20$ other density ramp discharges, with slight variation in timing of changes (e.g. drop in target density as determined by Langmuir probes) and magnitude (e.g. the total recombination).

The vertical error bars shown in figures \ref{fig:Overview50648}A-H represent 95 \% confidence intervals. Measurement uncertainties have been determined by propagating measurement uncertainties in the absolute calibration; in fit parameters (determined through the Gradient Expansion Method \cite{BEVINGTON2003}); in $\Delta L$ and assuming $n_o/n_e$ is in between 0.01 and 0.25. 



Bolometry and spectral features consistently indicate an expansion of a cold plasma region from the target towards the x-point during a density ramp. During a considerable increase in Greenwald fraction from $\overline{n_e}/n_G = 0.3$ to $\overline{n_e}/n_G = 0.5$ (figure \ref{fig:Overview50648}A, $F_{76}$ increases resulting in an increase in $F_{rec}$ from $<0.35$ to $\sim 1$ (figure \ref{fig:Overview50648}C). Significant increases in $F_{rec}$ first occur near the target and later the region of enhanced $F_{rec}$ expands towards the x-point. The radiation front as measured by bolometry (figure \ref{fig:Overview50648}E), which is representative of higher temperatures than those at which recombination occurs \cite{HARRISON2016PSI,STANGEBY2000}, also moves from the target towards the x-point and is correlated with the increase in $F_{rec}$. 


The above spectral features are consistent with a strong recombining region near the target. Those features include a strong increase in $B_{n\rightarrow2}$ (figure \ref{fig:Overview50648}B) which, combined with a rising $F_{rec} (n=7)$, implies that $R_L$ (figure \ref{fig:Overview50648}G) is strongly increasing. Similar to trends in $F_{rec}$, the onset of this non-linear increase starts first close to the target and later increases closer to the x-point. The increase in $B_{n \rightarrow 2}$ during the density ramp is both due to the $n_e$ increase (figure \ref{fig:Overview50648}D) and $T_e^{avg}$ decrease (figure \ref{fig:Overview50648}H).

Our results suggest that recombination is insufficient to effectively reduce the particle flux at the time of the particle flux roll-over. Furthermore, the (Stark) density close to the target does not decrease. After $F_{rec} \rightarrow 1$,  $B_{n \rightarrow 2}$, $R_L$ and $n_e^{Stark}$ keep increasing until the end of the discharge while remaining highest at the lowest DSS chord 5 cm above the target. At first glance this and bolometric measurements (figure \ref{fig:Overview50648}E) would seem to indicate that while ionization and impurity radiation have detached from the target, the high density region has not. However, Langmuir probe data (taken from the probe closest to the separatrix) suggests the density has dropped at the target (figure \ref{fig:Overview50648}) as discussed in section \ref{RecLP}. 

It is possible that the inferred $R_L$ is an underestimate, since the closest target DSS view line intersects the separatrix 5 cm above the strike point. If the $R_L$ spatial profile is extrapolated to the target, $R_L$ at the target is three times higher than at the DSS chord closest to the target. However, target probe measurements (figure \ref{fig:Overview50648}D) indicate $n_e$ drops in this non-observed region, which would lower $R_L$. Combining LP data and spectroscopic data (section \ref{RecLP}) suggests that either the electron temperature at the target is very low ($< 0.06$ eV) or the high-density recombination front has moved off-target and is located in the region between the target and the lowest DSS chord. Detachment in TCV has so far never reached the level where the density and recombination region peak moves to points above the lowest DSS chord.


The total recombination rate in the divertor $R_V$ [rec./s] is determined by integrating $R_L$ toroidally and poloidally across the chords. $R_V$ increases strongly during the last phase of the discharge (figure \ref{fig:Overview50648}F), and reaches values of up to $R_V = (6 \pm 2) \cdot 10^{21}$ rec/s, which is similar to the total particle flux measured by Langmuir probes at that time, indicating that $R_V$ contributes significantly to the particle flux drop at this time. However, the particle flux measured by the Langmuir probes drops at 1.0 s. $R_V$ at that time is relatively low, which indicates that recombination losses are not the main contributor to the initial particle flux drop. 

\subsection{Recombination signatures compared with Langmuir probe data}
\label{RecLP}

Combining data from the DSS and divertor target Langmuir probe data is informative about the development of detachment. The peak target electron density determined from Langmuir probe (LP) I-V characteristics is in agreement with $n_e^{Stark}$ near the target until 0.9 s (figure \ref{fig:Overview50648}D), which is close to the time when $B_{n\rightarrow2}$ starts increasing strongly. 

Across many tokamaks it has been found that the temperature derived from Langmuir probes is overestimated for $T_e < 5 \text{ eV}$, \cite{BATISHCHEV1997POP,BATISHCHEV1996POP}. Assuming this is also true for TCV, we utilize $J_{sat}$ and $n_e^{Stark}$ (5 cm from the target) to calculate $T_e^{\text{mod}}$ (figure \ref{fig:Overview50648}H). $T_e^{\text{mod}}$ decreases during the density ramp in agreement with the Balmer line derived $T_e^{avg}$ up until 1.1 s when both the target particle flux and target density (LP) have started dropping. Near the end of the discharge, $T_e^{mod}$ reaches temperatures below 0.06 eV, much lower than $T_e^{avg}$ obtained from Balmer line analysis ($\sim 0.5$ eV). In addition it should be noted that the peak density in the density profile along the line of sight  is likely higher (up to 35 \%) than the density inferred from Stark broadening due to the weighted average along the chord, which would lead to an even lower $T_e^{mod}$. Therefore, if $T_e$ at the target would be higher than 0.06 eV, it would imply the target density would be lower than $n_e^{Stark}$. Hence, the density front would have moved between the target and the first DSS chord. 




\section{Discussion}
\label{Disc}

The onset of detachment observed spectroscopically at TCV is generally similar to the dynamics previously observed at higher density machines, but there are also significant differences.

As the core density is increased in L-mode plasmas, the target density increases and the temperature decreases, which are general characteristics of a high-recycling divertor. However, the ion current to the target does not increase $\propto <n_e>^2$ as expected from the two point model (assuming $n_{e, up} \propto <n_e>$) \cite{STANGEBY2000} (figure \ref{fig:Overview50648}F). This difference to other, high density machines and the two-point model may be due to the fact that the ionization mean free path in TCV ($\lambda_{ioniz} \sim 5-10 \text{ cm}$) is larger compared to the width of the divertor plasma ($d_{fan} \sim$ a few cm) near the target \cite{WISCHMEIER2005EPFL}. Together with the open divertor geometry neutrals are not well-confined, which likely leads to less ionization and a slower rise in divertor density. That could reduce the amount of charge exchange and recombination, thus slowing down the detachment process.

Once the detachment process starts with the drop in divertor target density and the rise in recombination signatures ($F_{rec}$ and $R_L$, figures \ref{fig:Overview50648}C and \ref{fig:Overview50648}G) the process of detachment proceeds slowly in TCV. Instead of a swift movement of the recombination and high-density regions observed at other higher density machines \cite{LIPSCHULTZ1999POP, POTZEL2014NF}, the recombining region and peak density stays near the divertor target at TCV while recombination signatures extend towards the x-point. At the highest core and divertor densities in the TCV plasmas studied so far, the drop in target density (figure \ref{fig:Overview50648}D) concurrent with the continued increase in the DSS-inferred density is consistent with the detachment region (in the sense of both low density and low temperature) moving off the divertor target slightly, less than the 5 cm corresponding to the lowest DSS chord, as discussed in section \ref{RecLP}. However, such a movement is very slow given that the Stark-derived density continues to rise throughout the remainder of the discharge. 

The inference of recombination rates through the DSS data analysis also provide some insight into the role of recombination in removing ions from the plasma and causing a density drop. As discussed earlier, recombination remains highest near the target throughout the discharge, with the total amount of recombination rising rapidly to levels at the end of the discharge comparable to the target ion flux. Given that the target density drops earlier in the pulse and that the total recombination rate is less than 1\% of the particle flux at the point the particle flux starts dropping, the question is whether recombination is playing an important role at this time. The two possibilities are that the ion source upstream could start dropping at the same time as the target density falls. Or, that the recombination local to the flux tube of the peak ion flux is removing significant ion flux. We do not have enough spatial information at this time to comment further on the relative important of the two effects. 



\section{Summary}

The physics of the TCV divertor, including the detached regime, has been investigated at TCV, using a newly developed divertor spectroscopy system (DSS), together with advancements in techniques for extracting information from the Balmer spectrum. Analysis of the DSS data has been instrumental in characterizing the behaviour of detachment in TCV. We find that the detachment process develops slowly: the radiation first peaks near the divertor and then moves towards the x-point. The rise in the dominance of recombination signatures over excitation signatures follows the movement of the radiation peak, while the strongest level of density and recombination remains close to the target. Even as the plasma density above the target continues to increase the ion current to the target drops, which may imply that the detached (low pressure and density) region has moved off the target. But within the density range studied on TCV, the detachment front moves no further.

The role of recombination in ion loss has been investigated. We find that there is no clear connection between the initial roll-over in the target ion current and the level of recombination. However, later in time, $R_V$ approaches the integral ion current and it may be that the recombination front moves further off the target if higher densities could be achieved. Further studies are needed.  

\section{Acknowledgements}

This work has been carried out within the framework of the EUROfusion Consortium and has received funding from the Euratom research and training programme 2014-2018 under grant agreement No 633053. The views and opinions expressed herein do not necessarily reflect those of the European Commission.

\bibliographystyle{plainnat}
\bibliography{qiqqa07062016A}

\end{document}